\documentclass[9pt,twocolumn,twoside]{optica}
\setboolean{shortarticle}{true}
\setboolean{minireview}{false}

\usepackage[T1]{fontenc}       
\usepackage{color}
\usepackage{graphicx}
\usepackage{bm}
\usepackage{commath}
\usepackage{amsmath}
\usepackage{latexsym}
\usepackage{graphicx}
\usepackage{caption}
\usepackage{subcaption} 	 	
\usepackage{csquotes}
\usepackage{gensymb}
\usepackage{epstopdf}

\title{Highly Sensitive Wavelength-scale Amorphous Hybrid Plasmonic Detectors}

\author[1,*]{Yiwen Su}
\author[1]{Charles Lin}
\author[1]{PoHan Chang}
\author[1]{Amr S. Helmy}

\affil[1]{The Edward S. Rogers Sr. Department of Electrical and Computer Engineering, University of Toronto, 10 King's College Road, Toronto, ON M5S 3G4, Canada}

\affil[*]{Corresponding author: yw.su@mail.utoronto.ca}

\ociscodes{(040.0040) Detectors; (040.5160) Photodetectors; (040.6040) Silicon; (130.0130) Integrated Optics; (250.5403) Plasmonics; (310.6628) Subwavelength structures, nanostructures.}

\begin{abstract}
Hybrid integration of plasmonics and Si photonics is a promising architecture for global microprocessor interconnects. To this end, practical plasmonic devices not only should provide athermal, broadband operation over wavelength-scale footprint, but also support non-intrusive integration with low-loss Si waveguides as well as CMOS back-end-of-line processes. Here, we demonstrate a hybrid plasmonic photodetector with a single active junction fabricated via back-end deposited amorphous materials coupled to Si nanowires with only 1.5dB loss. Utilizing internal photoemission, our detectors measured sensitivity of -35dBm in a 620nm by 5$\mu$m footprint at 7V bias. Moreover, responsivity up to $0.4mA/W$ and dark current down to $0.2nA$ were obtained. The high process tolerance is demonstrated between $\lambda$=1.2-1.8$\mu m$ and up to $100\degree C$. The results suggest the potential towards plasmonic-photonic optoelectronic integration on top of Si chips without costly process modifications.
\end{abstract}

\setboolean{displaycopyright}{true}

\begin{document}
	
	\maketitle
	
	On-chip optical communications offer immense promise to enhance performance of emerging generations of microprocessors~\cite{Sun2014,Sun2015}. However, a realistic optoelectronic platform demands device miniaturization, athermal behavior, as well as broadband operation~\cite{RadamsonBook}. Many of the current guided-wave photodetector designs, such as ones utilizing 2D materials~\cite{Liu2013} or resonant cavities~\cite{Casalino2010,Haret2013}, cannot fulfill these requirements simultaneously. Although plasmonic designs show promising performance, they exhibit significant loss and photogeneration typically takes place in junctions implemented with crystalline materials inherently incompatible with CMOS back-end processing and thus dictate significant modifications to front-end manufacturing.~\cite{Goykhman2014,Casalino2013,BeriniSPP1,BeriniSPP2,silicide,Muehlbrandt}. Here, we report the first experimental demonstration of a wavelength-scale, non-resonant, high-sensitivity photodetector with a single active junction formed with amorphous materials. Employing internal photoemission (IPE) within a 5$\mu m$-long hybrid plasmonic waveguide, we obtained minimum sensitivity, dark current, and static power of $-35dBm$, 0.2$nA$, and 1.2$nW$ at 7V respectively. The device is tested to be operational between $15-100\degree C$ and $\lambda = 1.2-1.8\mu m$. Moreover, with only 1.5 dB coupling loss to Si nanowires, our design demonstrates the potential for integrating amorphous-based plasmonic devices and low-loss dielectric waveguide interconnects to form a densely-packed, low-loss optoelectronic platform.
	
	\begin{figure*}[!t]
		\centering
		\includegraphics[width=0.78\textwidth]{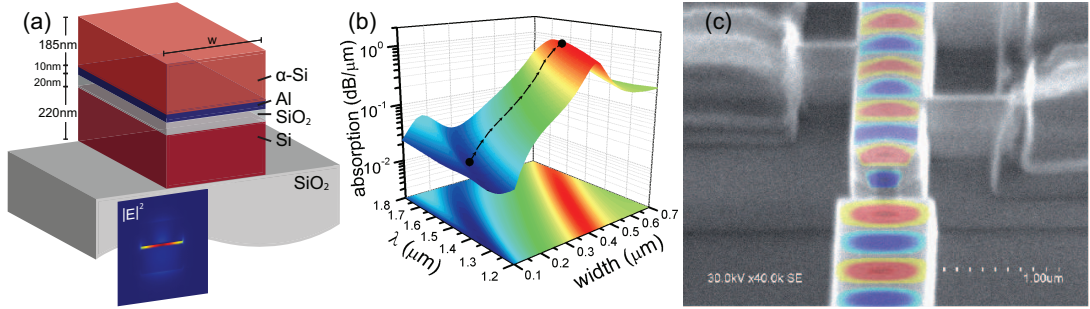}
		\caption{(a) Schematic of the AHPW and field profile of its symmetric supermode. Light is primarily confined at the $\alpha$Si-Al interface, leading to enhanced light-matter interaction for IPE at the $\alpha$Si-Al interface. (b) Waveguide absorption as a function of waveguide width and wavelength. The waveguide platform is versatile as it is capable of supporting low insertion loss (width of $200nm$) as well as significant absorption (width of $620nm$) under the same fabrication process. (c) Direct endfire excitation of the AHPW via TM-polarized light from a Si nanowire with $220nm$-thickness and $800nm$-width with a coupling loss of only $1.5dB$.}
		\label{fig:modes}
	\end{figure*}
	
	\begin{figure}[!th]
		\centering
		\includegraphics[width=0.7\linewidth]{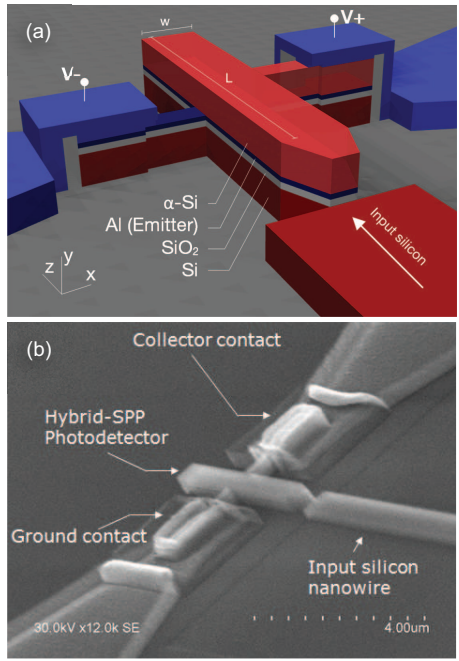}
		\caption{(a) Schematic and (b) SEM of the AHPW photodetector. The layer thicknesses are indicated in Fig. 1a. A $500nm$-long taper is utilized to link between the AHPW coupler and photodetector sections, which have width of $200nm$ and $620nm$ respectively. The cut-through view of contact fingers shows two reverse biased Schottky junctions in MSM configuration separated by the $\alpha$Si layer. Micron-sized collector contacts are deposited on fingers extending away from the AHPW core to minimize scattering of the optical mode.}
		\label{fig:physical}
	\end{figure}
	
	The photodetector reported is based on asymmetric hybrid plasmonic waveguides (AHPWs), consisting of Si-$\mathrm{SiO_{2}}$-Al-$\alpha$Si layers (Fig.~\ref{fig:modes}a). In this platform, the hybrid plasmonic waveguide (HPW) mode supported by the Al-$\mathrm{SiO_{2}}$-Si layers and the surface plasmon polariton (SPP) mode supported by the $\alpha$Si-Al layers are coupled to form supermodes. Specifically, the symmetric supermode, which corresponds to out-of-phase coupling of the evanescent fields at the metal layer, is utilized as the signal carrier in our design because of its highly tunable absorption characteristics~\cite{Oulton2008,AlamCLEO2007,WenAHPW}. The theory and operation principle of the AHPW are discussed in the Suppl. Mat.
	
	The loss characteristics of AHPW supermodes are highly sensitive to the symmetry of the evanescent fields that are superimposed across the common metal layer. This symmetry can be controlled by manipulating structural parameters such as core width (Fig.~\ref{fig:modes}b). Specifically, complete field symmetry is established for waveguide width of 200$nm$ and the waveguide loss is reduced to only $0.02dB/\mu m$, an order of magnitude lower than either that of the HPW and SPP-waveguide ($0.1dB/\mu m$ and $0.45dB/\mu m$ respectively). Moreover, small momentum and field mismatch between the symmetric supermode of AHPW and TM mode of a Si nanowire allows the $200nm$ AHPW to be excited from an $800nm$ Si nanowire via non-resonant butt-coupling with only $1.5dB$ coupling loss (Fig.~\ref{fig:modes}c).

	\begin{figure}[!t]
		\centering
		\includegraphics[width=0.8\linewidth]{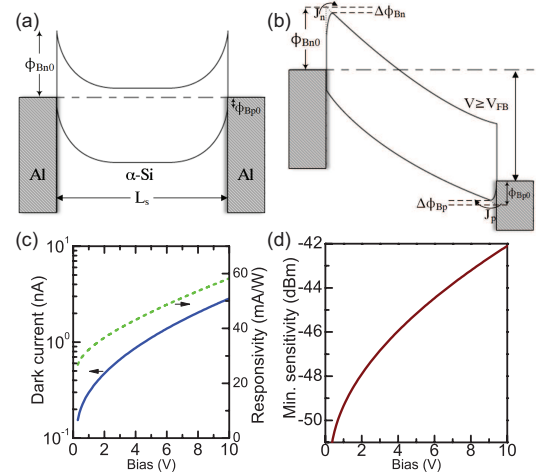}
		\caption{(a) Energy band diagram for an active MSM junction with $\alpha$Si gap ($L_{s}$) of $185nm$ when not biased, where $\Phi_{Bn0}$ = 0.54 eV and $\Phi_{Bp0}$ = 0.58 eV are the intrinsic electron and hole barrier heights. (b) Energy band diagram for the same junction when biased at or above flatband voltage ($V_{FB}$), computed to be 0.3V for a moderate doping concentration of $10^{16} cm^{-3}$. $\Delta\Phi_{Bn}$ and $\Delta\Phi_{Bp}$ are the barrier height reductions due to image force lowering effect at $V > V_{FB}$. $J_n$ and $J_p$ are the dark current density contributions from electron and hole injection respectively. (c) Computed dark current, responsivity, and (d) minimum sensitivity for a 10$\mu m$ long AHPW photodetector with 90\% absorption. See Suppl. Mat. for detailed modeling.}
		\label{fig:operation}
	\end{figure}

	As waveguide width increases, the absorption of the symmetric supermode increases exponentially due to combined effects of field symmetry breaking and modal evolution asymmetry (see Suppl. Mat.). For $\lambda = 1.55\mu m$, the absorption loss peaks at $1.0dB/\mu m$ at a width of $620nm$. Thus, without modifying the vertical dimensions or materials, an AHPW can be engineered to serve either as a long-range propagating passive component that interfaces efficiently with silicon photonics or a highly absorptive active device such as photodetector. This is achieved with only a 10$nm$-thin metal layer, thus heavily alleviating the loss of photocarrier energy due to scattering in the metal.
	
	Fig.~\ref{fig:physical}a illustrates the schematic and biasing of a AHPW travelling-wave photodetector that has CMOS-compatible amorphous Si top layer. A $500nm$-long taper is utilized to link between the AHPW coupler and photodetector sections, which have width of $200nm$ and $620nm$ respectively. The photodetection is based on IPE (see Suppl. Mat. for details of the mechanism and model). The $\alpha$Si-Al interface is utilized to support both the SPP portion of the AHPW and to serve as the Schottky junction emitter for the IPE mechanism. With the addition of an Al collector on the opposite side of the $\alpha$Si layer, the junctions form a metal-semiconductor-metal (MSM) device in the vertical direction and consist of two back-to-back Schottky junctions separated by the $\alpha$Si thickness~\cite{Sze1971}. Micron-sized collector contacts are deposited away from the AHPW core to minimize the junction area and hence dark currents, as well as reduce optical mode scattering.
	
	The energy band diagrams of the Al-$\alpha$Si-Al junction at unbiased thermal equilibrium state and at flatband voltage are shown in Fig.~\ref{fig:operation}a and~\ref{fig:operation}b respectively. The Schottky barrier heights for Al/Si junctions are assumed to be $\Phi_{Bn0} = 0.54eV$ and $\Phi_{Bp0} = 0.58eV$~\cite{SzeBook}. As the SPP mode propagates, energy is dissipated and absorbed into the lossy Al layer along the waveguide. The absorbed photons with energy $\mathrm{h}\nu$ higher than the barriers height have a probability to cross into the $\alpha$Si layer. In the unbiased case, no appreciable current flow can be generated due to the built-in potential of the collector junction impeding photogenerated carriers to cross to the opposite terminal. As external bias is increased to the flatband condition, the depletion width extends throughout the thickness of the $\alpha$Si layer, allowing photocurrent carriers to be swept rapidly from the emitter to the collector. Applying further bias reduces the effective barrier height due to image force lowering effect, which increases both detector responsivity and dark currents.
	
	The responsivity, dark current, and minimum sensitivity of a 10$\mu m$-long AHPW photodetector with 90\% absorption are plotted in Fig.~\ref{fig:operation}c and~\ref{fig:operation}d. The collector contact junction areas are designed to be $1\times1\mu m^2$ and in order to utilize the lower n-barrier for carrier emission, the 10$nm$-thick Al emitter is biased at a negative potential with respect to the collector on top. With an absorption of $1.0dB/\mu m$ for the symmetric supermode, responsivity of 30$mA/W$ can be achieved under 2$V$. The dark current density is on the order of 1$nA/\mu m^2$ and the minimum theoretical sensitivity of the detector can reach $-50dBm$.
	
	\begin{figure}[!t]
		\centering
		\includegraphics[width=0.8\linewidth]{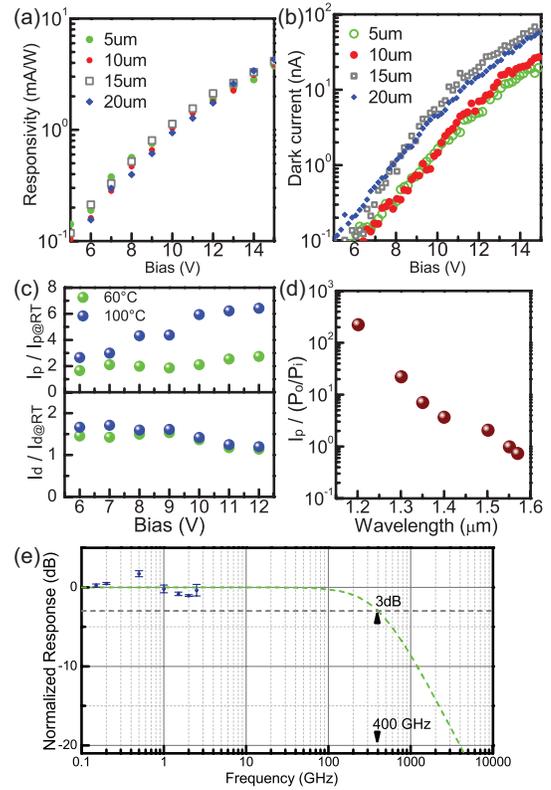}
		\caption{(a) Measured responsivities and (b) dark currents for photodetector lengths of $5-20\mu m$ at room temperature. (c) Measured photocurrents and dark currents for 10$\mu m$-long detectors at $60\degree C$ and $100\degree C$, normalized with respect to room temperature. The sensitivity improves with temperature as the increase in photogeneration is more dominant than dark currents. (d) Measured photocurrents as a function of wavelength, normalized to photogeneration at $\lambda = 1550nm$ and the coupling efficiency from a Si nanowire. The device is only tested up to $1.575\mu m$ due to equipment limitations but is expected to be operational up to $1.8\mu m$ based on simulation. (e) Frequency response measured between $100MHz$ and $2.5GHz$ fitted to the theoretical RC response, modeled using a $50\Omega$ load resistance and $8fF$ measured capacitance. However, the upper limit in speed of the current generation of devices is expected to be dictated by carrier transit time.}	
		\label{fig:pd_results}
	\end{figure}
	
	The fabricated AHPW photodetector is shown in Fig.~\ref{fig:physical}b. The bottom Si layer is crystalline since standard Si-on-insulator wafers were utilized for proof-of-concept. However, the active junction is entirely implemented via sputtered Al and $\alpha$Si.  Using the cut-back method, the coupling efficiency from an input Si nanowire into the AHPW photodetector is measured to be 70\%. The measured responsivities and dark currents at $\lambda$ = 1550 nm are plotted in Fig.~\ref{fig:pd_results}a and~\ref{fig:pd_results}b for detector lengths of $5-20\mu m$. While the metrics obtained from simulations assumed a 10$\mu m$-long device for 90\% absorption, in our experiments it was found that a $5\mu m$ photodetector can report responsivity as high as 5.0$mA/W$ in the tested voltage bias ranging from 6 to 15$V$ at room temperature. The data for below 5V are not shown since the measured current falls below the detectable level of the instruments. However, the lower responsivity compared to theory can be attributed to the recombination of photocarriers in dangling bonds of the $\alpha$Si, which was not accounted for in our model, and insufficient doping of the $\alpha$Si, as evident in the high voltage required to observe photodetection at biases above $6V$ and lack of a sharp flatband response. However, when evaluating the photodetector using minimum sensitivity, a metric that accounts for the trade-off between detector responsivity and dark current and measures the minimum optical power required for detection, we achieve $-32dBm$ at room temperature due to extremely low dark currents down to 0.2$nA$. The quiescent power consumption of the device is therefore 1.2$nW$. Similar responsitivies are obtained for longer detector lengths, but dark currents become significantly higher due to thermionic emission from increased active junction areas. This suggests that the plasmonic mode is highly absorptive and coupled light is absorbed within the first few microns. Note that dark current for longer junctions are similar, attributed to non-uniform formation of Schottky junctions which are known to be difficult to control, as well as contact pad resistance changes due to probing damage.
	
	The photodetector performance at elevated temperature is displayed in Fig.~\ref{fig:pd_results}c. The devices have been tested between $15-100\degree C$, the maximum range of our instruments. While an increase in dark currents typical of thermionic emission is observed, the devices also display an increase in photocurrents as a result of the Schottky barrier height reduction in $\alpha$Si as temperature increases~\cite{Bapat1987}. It is observed that dark currents increase by 50\% as temperatures are increased to $100\degree C$ while photocurrents can increase by factors of 2 to 3 at low bias operation. Thus, the effects of barrier height reduction are more dominant for photocarrier generation in the $1550nm$ wavelength regime than it is for dark currents. Overall, the devices are fully functional up to $100\degree C$ and the sensitivity can be improved to $-35dBm$ as temperature increases.
	
	The optical bandwidth of the device is characterized from $1.1\mu m$ to $1.6 \mu m$ (Fig.~\ref{fig:pd_results}d). The photocurrents are normalized to the optical power coupling efficiency extracted from reference passive AHPW transmission measurements. It is observed that while the optical performance of the AHPW is optimized for maximal coupling and operation at $\lambda = 1550nm$, normalized responsivity can increase as wavelength decreases. As the wavelength shortens to $1.35\mu m$, the decreased optical power coupled into the device is compensated by higher internal quantum efficiency as photon energies increase over the Schottky barrier height and still operates mainly on IPE. As the wavelength is further decreased, it is suspected that the main detection mechanism becomes direct absorption from the band-tail of $\alpha$Si, which increases up to two orders of magnitude higher than what is achieved through IPE. Finally, in the regime below $1.2\mu m$, the input $c$Si nanowires become absorptive and the optical power transmission falls below the detectable levels of the characterization setup. Based on theoretical simulations, the photodetector is expected to operate up to 1.8$\mu m$, at which the optical mode becomes leaky (see Suppl. Mat.).
	
	The frequency response, shown in Fig.~\ref{fig:pd_results}e, was measured in the frequency range between $100MHz$ and $2.5GHz$, dictated by the available instrumentation. On the same plot, the RC bandwidth was plotted using the values measured for these devices. The capacitance was found to be in the range of $8-10 fF$ depending on device length (See Suppl. Mat). The RC bandwidth is calculated to be $400GHz$ for a $50\Omega$ load. However, speed of the current generation is expected to be limited by carrier transit time governed by the saturation drift velocity of $\alpha$Si due to poor electrical properties~\cite{Maassen2007}. Future generations will alleviate this limitation through controlled doping profiles as well as reducing the $\alpha$Si layer thickness to reduce transit time.
		
	The detailed comparison between existing IPE-based plasmonic photodetector designs can be found in Suppl. Mat. To the best of our knowledge, this work is the first demonstration of a hybrid plasmonic detector architecture. Moreover, these specifications are achieved via amorphous materials for the first time. Despite the use of Al instead of more photogeneration efficient, but non-CMOS-compatible layers such as Au, our device boasts $-35dBm$ minimum sensitivity. In comparison, the closest comparable plasmonic architecture based on metal strip waveguides on crystalline Si only measures $-14dBm$ with devices that are 8 times longer in length~\cite{BeriniSPP1,BeriniSPP2}. In plasmonic photodetectors employing metal silicides on crystalline silicon, similar sensitivity values are achieved ($-30dBm$) on devices with lengths of $20\mu m$, but formation of silicide junctions requires high-temperature fabrication techniques incompatible with back-end processing~\cite{silicide}.
	
	In conclusion, we reported the first experimental demonstration of an integrated, hybrid plasmonic photodetector with an active region based on deposited, amorphous materials. The use of non-CMOS-compatible Au or crystalline Si can further improve the device responsivity~\cite{Muehlbrandt}; however, our use of Al and amorphous Si still provides $-35dBm$ minimum sensitivity, the best reported up to date. This is enabled by the nanoscale mode area thus enhanced light-matter interaction in the AHPW, allowing for efficient photogeneration and reduction of device length, dark currents and power consumption. The photodetectors were shown to be operational between $1.2-1.8\mu m$ and functional at temperatures tested up to $100\degree C$. While our prototype devices were built on a Si-on-insulator platform, the purpose was to demonstrate the highly efficient power coupling from Si nanowires and fairly compare against designs in literature that are based on similar integration schemes with Si photonics. The photogeneration process is based on IPE and takes place entirely in the active junction formed from the metal and an amorphous material while still showing much improved performance compared to previous designs based on crystalline Si. The use of amorphous materials and the efficient integration of our hybrid plasmonic architecture to Si photonics opens up new frontiers in integrated optoelectronics as they allow for non-intrusive CMOS back-end integration with low-loss dielectric waveguide interconnects on microprocessors away from noisy transistor levels and expensive modifications of front-end-of-line CMOS manufacturing.
	
	
	\section*{Supplemental Documents}
	
	\noindent See \href{link}{Supplement 1} for supporting content.

\end{document}